\documentclass[useAMS,usenatbib]{mn2e}
\usepackage{gensymb}
\usepackage{hyperref}
\usepackage{graphicx}
\usepackage{float}
\usepackage{multicol}
\usepackage{multirow}
\usepackage{longtable}
\usepackage{lipsum}
\usepackage{relsize}
\title[Comet C/2014 Q2 (Lovejoy)]{Optical Spectroscopy of Comet C/2014 Q2 (Lovejoy) from MIRO}
\author[Kumar et al.]{Kumar Venkataramani $^{1,2}$\thanks{E-mail : kumar@prl.res.in}, Satyesh Ghetiya$^{1,3}$, Shashikiran Ganesh$^{1}$\thanks{E-mail: shashi@prl.res.in}, \and U.C.Joshi$^{1}$, Vikrant K Agnihotri$^{4}$, K.S.Baliyan$^{1}$ 
\\ \\
$^{1}$Physical Research Laboratory, Ahmedabad, India\\
$^{2}$Indian Institute of Technology Gandhinagar \\
$^{3}$Indian Institute of Space Science and Technology, Thiruvananthapuram, India \\
$^{4}$Cepheids Astronomy, Kota, India}

\begin{document}

\date{Accepted for publication in MNRAS}


\pagerange{\pageref{firstpage}--\pageref{lastpage}} \pubyear{2002}

\maketitle

\label{firstpage}

\begin{abstract}

Spectra of comet C/2014 Q2 (Lovejoy) were taken with a low resolution spectrograph mounted on the 0.5 m telescope at the Mount Abu Infrared Observatory (MIRO), India during January to May 2015 covering the perihelion and post-perihelion periods. The spectra showed strong molecular emission bands (C$_{2}$, C$_{3}$ and CN) in January, close to perihelion. We have obtained the scale lengths for these molecules by fitting the Haser model to the observed column densities. The variation of gas production rates and production rate ratios with heliocentric distance were studied. The extent of the dust continuum using the $Af\rho$ parameter and its variation with the heliocentric distance were also investigated. The comet is seen to become more active in the post-perihelion phase, thereby showing an asymmetric behaviour about the perihelion.

\end{abstract}

\begin{keywords}

comet:general-comet:individual: C/2014 Q2 - methods: observational - techniques: spectroscopic - telescopes - Oort Cloud

\end{keywords}

\section{Introduction}

Comets are cold icy bodies in the solar system that were formed in the solar nebula and are considered to be the signature bodies to understand the formation of the solar system. As the comet nucleus makes its journey towards its parent star, the ices start sublimating giving rise to a mixture of gas and dust which form the coma. For comets at heliocentric distances less than 3 AU, the visible band spectrum shows strong molecular emission bands riding on the continuum radiation scattered by the cometary dust. Studying these molecular emission bands has been an important part of the cometary study. \\ Comet C/2014 Q2 (Lovejoy), an Oort cloud comet, was discovered by Terry Lovejoy in August 2014 using an 8 inch telescope when the comet had a visual magnitude of 15. The comet brightened to a magnitude of 4 in the visual band in January 2015 being closest to Earth at 0.469 AU on January 7. The comet's perihelion was on the 30th of January 2015 at a heliocentric distance of 1.29 AU. According to the data from JPL Horizons database, the comet orbit has an eccentricity of 0.9976, orbital inclination of 80.3$^{\degree}$, semi-major axis of 576.34 AU and orbital period of 13700 years.  \\
\citet{b1} have carried out a detailed survey of 85 comets during the years 1976-1992. They have studied the variation of gas production rates and dust to gas ratio with heliocentric distances. They have defined certain limits in the gas ratios in order to classify the comets as to whether they are depleted in carbon-chain molecules or not. According to their study, most of the carbon depleted comets are from the Jupiter family. They also say that CN is produced from the nucleus as well as from the dust in the comet's coma. Several spectroscopic surveys have been carried out, e.g. \citet{Newburn}, \citet{Fink_2009}, \citet{b2}. \citet{b2} have spectroscopically studied five comets. They found a linear correlation between production rates of C$_{2}(\Delta \nu=0)$, C$_{2}(\Delta \nu=1)$ and C$_{3}$ with respect to CN. No correlation was found between the production rate ratios and heliocentric distance.  \citet{Langland} have given new statistical methods in cometary spectroscopy for the extraction and subtraction of the sky using the comet frames itself. They have given a dynamical classification of comets by studying 26 different comets. More recently, \citet{Cochran2012} have compiled and reported the spectroscopic results for 130 comets observed from McDonald Observatory. They have found remarkable similarity in the composition of most of the comets. They quote that the carbon chain depleted comets can be from any dynamical class. However they have not found any of them from the Halley type comets (Tisserand parameter $<$ 2 and period $<$ 200 years) due to significantly low number of these types in their sample. \\ In our study of comet C/2014 Q2,  we have determined the production rates for various gas species, $Af\rho$ and dust to gas ratios at different heliocentric distances.
\section{Observations}
The observations were carried out with LISA spectrograph mounted on the 0.5 m(f/6.8) telescope (PlaneWave Instruments CDK20) at the Mount Abu Infra-red observatory (MIRO), Mount Abu, India. \citet{Ganesh} have given a brief overview of the telescope at the observatory. In the following we describe, briefly, the spectrograph and comet observations.
\subsection{Instrument for observation : The LISA spectrograph}
\label{LISA}
LISA\footnotemark[1] is a low resolution high luminosity spectrograph which is designed for the spectroscopic study of faint and extended objects. The light from the object that passes through the slit falls on the grating. The rest of the light is reflected towards the guiding CCD camera. Two detector cameras are placed on either side of the spectrograph : Atik 314L is used for acquiring the object spectrum. This has a chip size of 1392 X 1040 pixels where each is a square pixel of size 6.45 $\mu m$. The light that passes through the slit is reflected by the mirror and is transformed into a parallel beam by a collimator. This is then diffracted by the grating and then passes through the objective lens and gets converged onto the detector. As the focal length of the collimator is 130 mm and that of the camera lens is 88 mm, the image on the slit plane is multiplied by a factor of 88/130=0.68 onto the plane of the CCD. The second detector is Atik Titan guiding camera. This has a chip size of 659 x 494 pixels where each is a square pixel of size 7.4 $\mu m$. All the reflected light that does not pass through the slit is directed to the guiding mirror. This is then sent to the guiding optics which refocuses the light on to the guiding camera.
There are 8 different slits that are provided with the instrument. We have used a slit width of 1.37 arcsec for January and February, and a slit width of 2.08 arcsec for March and May observations. Grating with 300 grooves/mm was used covering a total wavelength range from $\scriptstyle\mathtt{\sim}$ 3800 $\AA$ to $\scriptstyle\mathtt{\sim}$ 7400 $\AA$ with a wavelength scale of 2.6 $\AA$ per pixel. The plate scale on the CCD plane is 0.57 arcsec per pixel. The length of the slit imaged onto the CCD plane is 130 arcsec.
\begin{figure*}
  \includegraphics[width=0.95\textwidth]{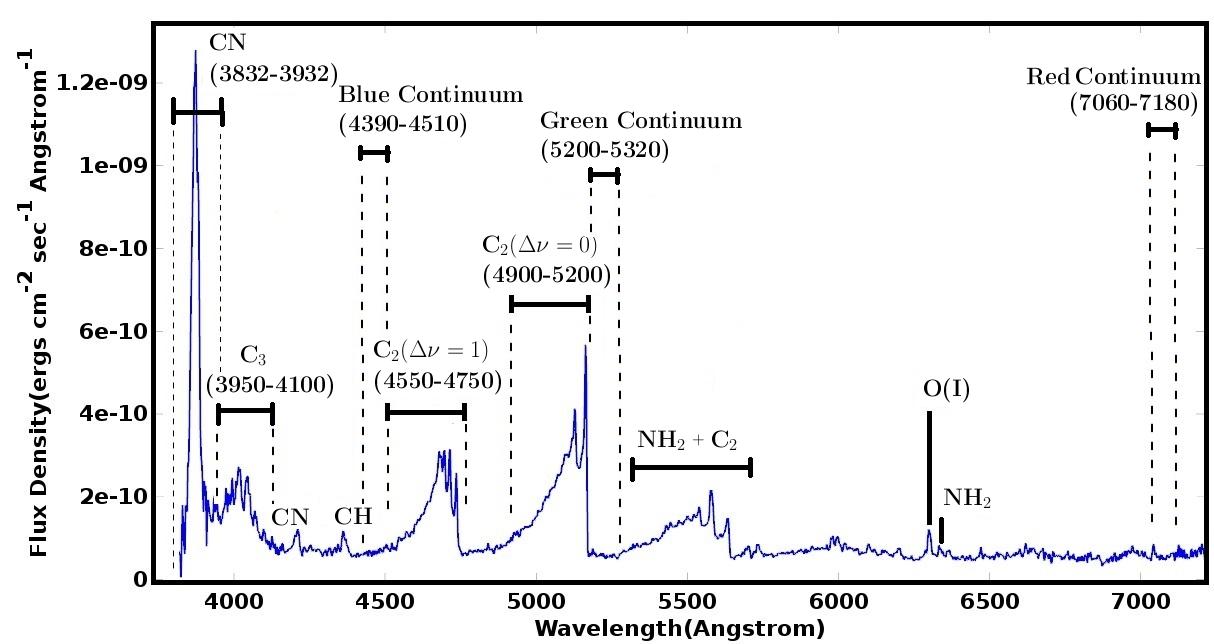}
 \caption{Flux Calibrated spectrum of Comet C/2014 Q2 (Lovejoy) : 1.37 $\times$ 163 arcsec aperture}
 \label{spectra}
\end{figure*}
\footnotetext[1]{More details on the spectrograph are available on the web site of the manufacturer: Shelyak Instruments (\url{http://www.shelyak.com/rubrique.php?id_rubrique=12&lang=2})}

\subsection{Comet observations}
\vspace*{-0.1cm} We carried out spectroscopic observations from Mount Abu Infra-Red Observatory (MIRO) using LISA, a low resolution spectrograph mounted on the 0.5 m telescope  during January to May 2015. The sky conditions were photometric during the observing period. Bias images were taken at regular intervals. The slit was assumed to be uniformly illuminated. With this assumption, lamp flats were taken after every round of observation. This was used for flat field corrections. An Argon-Neon lamp was used for the wavelength calibration. Details of the comet observations are given in table \ref{comet_obs}. The exposure times mentioned in the table are for each individual frame. The slit was placed at the photo-center of the comet and was manually tracked through the guiding CCD throughout  the exposure time. The observations were made using the scheme, sky-object-sky, for the proper background sky subtraction. The standard stars (along with their spectral types) from IRAF's catalogue of spectroscopic standards\footnotemark[2] observed for flux calibration are  as follows : HD15318(B9), HD19445 (G2V), HR1544(A1), HD126991(G2V), HD129357(G2V), HD140514(G2V), HR7596(A0), HD192281(O4.5), HD217086(O7). These were preferably observed at a similar airmass as that of the comet.

\begin{table}
\caption{Observational log. Columns: Date, Mid-UT, heliocentric distance in AU, geocentric distance in AU, Phase Angle in degree, Exp.(Exposure time) in second, Airmass}
\begin{tabular}{|c | c  | c | c | c | c | c |}
\hline 
Date & UT & r & $\Delta$ & Phase & Exp. & Airmass \\
\hline
27/01/2015 & 16:49 & 1.29 & 0.69 & 48 & 300 & 1.7 \\

23/02/2015 & 14:08 & 1.34 & 1.20 & 45 & 300 & 1.5\\

23/03/2015 & 14:45 & 1.50 & 1.66 & 36 & 600 & 2.8 \\

18/05/2015 & 22:20 & 2.03 & 2.21 & 27 & 1200 & 2.4 \\

19/05/2015 & 20:25 & 2.03 & 2.22 & 27 & 1200 & 2.7 \\

21/05/2015 & 22:24 & 2.05 & 2.23 & 27 & 1200 & 2.3 \\ 

\hline
\end{tabular}
\label{comet_obs}
\end{table}

\footnotetext[2]{The full list of spectroscopic standards available in IRAF can be found in (\url{http://stsdas.stsci.edu/cgi-bin/gethelp.cgi?onedstds})}

\section{Data Reduction and Analysis}
Data were reduced using IRAF, following the standard procedures of bias subtraction from all the images. The flat-field images were examined and the non-exposed part of the CCD image frame was trimmed in order to cut-off the sharp edges of the slit. The flats were median combined to form a master-flat. The wavelength dependence of the flat field images were corrected by normalizing the master-flat along the dispersion axis. This was then used for flat-fielding the object and the standard star images. As long exposures were taken, the comet spectrum and the background sky spectrum had to be corrected for the cosmic ray hits. Once this was done, the background sky image was subtracted from the comet image. Although this results only in partial sky subtraction, majority of the sky lines were removed by this process. IRAF's APALL task was used to extract the spectra of the comet, calibration lamp and standard stars. Using this task, different apertures can be set along the spatial axis and can be traced along the dispersion axis in order to extract the spectrum. The Ne-Ar lines were identified and were used to calibrate the wavelength axis of the comet and standard star spectra. The observed standard star spectra were compared with available flux values in the catalogue and an instrument sensitivity function was generated for each night. The airmass and extinction corrections were done by fitting an extinction curve. These were then used for the flux calibration of the comet spectrum. A flux calibrated spectrum of comet C/2014 Q2 (Lovejoy), taken on 27th January 2015 using LISA at MIRO is shown in figure \ref{spectra}. \\
In most of the cases, the seeing is larger than the slit width and therefore some of the light from the standard star does not enter the slit. Since the same slit is used for obtaining the comet and the standard star spectra, the cometary flux is overestimated during the flux calibration process. Appropriate correction factor is included in order to account for this overestimation. Corrections were also made in order to account for the differential refraction effect which leads to incorrect flux estimation at wavelengths, other than that used for guiding.   
\subsection{Gas Production rate}
The flux values from different emission lines were converted to molecular production rates which were calculated using the expression 
\begin{equation}
Q=\frac{4\pi \Delta^{2}}{g \ \tau} \times Flux \times Haser \ Correction 
\label{Q_eq}
\end{equation}
where $\Delta$ is the earth-comet distance, $g$ is the fluorescence efficiency of the molecule and $\tau$ is the lifetime of the molecule. The flux in equation (\ref{Q_eq}) represents the total flux obtained using an aperture of radius $\rho$. Using the Haser coma outflow model, \citet{Fink_1996} define the Haser correction as the inverse of a fraction. This fraction, called as the Haser fraction\footnotemark[3], is the ratio of number of emitting species within an arbitrary aperture to the total number of emitting species, if the aperture were to be extended to infinity. In order to do this, the total flux extracted from a circular aperture of radius $\rho$ needs to be known. Because we are doing long slit spectroscopy, we need to convert the long slit aperture into an equivalent circular aperture. For this purpose, we extract the flux using small apertures at various distances along the slit length and obtain the flux per unit area for each of them. We then multiply by the area of an annulus at that distance from the photo-center. The sum of the fluxes obtained in the concentric annuli will give the total flux from the circular aperture. 
Since fluorescence process gives rise to most of the emission lines in the visible region, the production rate of a molecule will depend upon the fluorescence efficiency of the molecule. This in turn depends upon the amount of solar flux that is received, the absorption oscillator strength and the relative downward transition probabilities for the molecule. The g-factor values for C$_{2}$ and C$_{3}$ are taken from \citet{b1}. The fluorescence efficiency of CN molecule depends on the heliocentric distance as well as the heliocentric velocity. \citet{tatum} has given the g-factor values for the CN molecule, taking into account, the Swings resonance-fluorescence effect. More recently, \citet{CN_Schleicher} has tabulated the g-factor values for various heliocentric distances and velocities, which we have used in our calculations. A double interpolation of the values in Table 2 of \citet{CN_Schleicher} is performed to get g value at the desired distance and velocity\footnotemark[3]\footnotetext[3]{Calculations using David Schleicher's web page (including Haser fractions) (\url{http://asteroid.lowell.edu/comet/comet_intro.html}).}. The scale lengths are taken from \citet{b1}, considering these species to be the daughter products. The scale lengths are scaled by r$_{h}^{2}$ and the fluorescence efficiencies are scaled by r$_{h}^{-2}$ in order to get the appropriate values at their respective heliocentric distances(r$_{h}$). The g-factors, scale lengths and Haser fractions used in the calculations have been listed in table \ref{gsh}. 
The continuum points defined by the HB narrow band filters, IHW filters and few other points were used to fit a polynomial function which was then used to obtain the continua of the spectra. In order to calculate the total integrated flux from the emission bands, the fitted continuum function was subtracted from the comet spectra to get clean continuum subtracted emission line spectra. The integrated flux values obtained for the emission bands CN(3832-3932 $\AA$), C$_{2}$(4900-5200 $\AA$), C$_{2}$(4550-4750 $\AA$) and C$_{3}$(3950-4100 $\AA$) were used to calculate the production rates of the corresponding molecule.

\begin{table*}
\caption{Scale lengths ($\beta$) in km, g-factor in ergs s$^{-1}$ mol$^{-1}$ and Haser fractions(H) for different molecular species at different heliocentric distances$^{\dagger}$. }
\begin{tabular}{c | c  | c | c | c | c |c| c | c|c|c|c|c|c|c|}
\hline \\
     &      &          &            &   \multicolumn{3}{|c|}{CN(3832-3932 $\AA$ )} &\multicolumn{3}{|c|}{C$_{2}$(4900-5200 $\AA$ )} &\multicolumn{3}{|c|}{C$_{3}$ (3950-4100 $\AA$ )} \\ 
 \cline{5-7} \cline{8-10} \cline{11-13} \\
Date & r    & $\Delta$ & log $\rho$ & g & $\beta$ & H &  g & $\beta$ & H & g & $\beta$ & H\\ 
     & (AU) & (AU) & (km) & $\times 10^{-13}$ & $\times 10^{5}$  & $\times 10^{-3}$ & $\times 10^{-13}$ & $\times 10^{5}$  & $\times 10^{-3}$ & $\times 10^{-13}$ & $\times 10^{5}$  & $\times 10^{-2}$ &  \\
\hline 
27/01/2015 & 1.29 & 0.692 & 4.60 & 1.6  & 3.5  & 3.6  & 2.7  & 1.1   & 7.2  & 6.0  & 0.4   & 6.5  & \\

23/02/2015 & 1.34 & 1.206 & 4.84 & 2.6   & 3.8  & 3.2  & 2.5   & 1.2 & 6.3 & 5.6  & 0.5 & 5.8 & &\\ 

23/03/2015 & 1.50 & 1.665 & 4.45 & 1.8   & 4.7   & 3.1   & 2.0 & 1.5 & 6.1 & 4.4  & 0.6 & 5.7 & & \\

19/05/2015 & 2.03 & 2.22 & 4.34 & 0.9 & 8.8  & 1.0  & 1.1  & 2.8 & 2.1 & 2.4  & 1.1 & 2.2 & & \\
\hline
\end{tabular}
\hspace*{-0.20in} $^{\dagger}$\small{All the values are given at their respective heliocentric distances}
\label{gsh}
\end{table*}

\subsubsection{Haser Model Estimation}
\label{Haser}
The Haser model is the simplest model to describe the outflow of gases in a comet. It is a model which describes the abundance of a molecular species based on a two step dissociation process. The model assumes a spherically symmetric coma of the comet and a uniform outflow of gas. \citet{KS} describes the Haser model in more detail. We have used the following equations as mentioned by \citet{Langland} in order to fit the Haser model to our observed gas distribution. The column density is calculated as 
\begin{equation}
N(y)=\frac{Q}{2\pi v_{flow}}\frac{\beta_{0}}{\beta_{1}-\beta_{0}} A(y)
\label{CD}
\end{equation}
where Q is the production rate in molecules per second, $v_{flow}$ is the outflow velocity in km per second, $\beta_{0}$ and $\beta_{1}$ are inverse of parent and daughter scale lengths in km respectively. $A(y)$ is given as 
\begin{equation}
 A(y)=\int^{z}_{-z} \frac{1}{y^{2}+z^{2}}\left(e^{-\beta_{0}\sqrt{y^{2}+z^{2}}}-e^{-\beta_{1}\sqrt{y^{2}+z^{2}}}\right)dz
\end{equation}
where $y$ is the projected distance from the center of the nucleus and $z$ is in the direction of line of sight. We assume that the comet's coma is spherically symmetric and therefore $2z$ is length of the cord at a distance y from the center. We extracted the spectrum using apertures of equal width at different distances $y'$ from the center. These were flux calibrated and the column density was calculated as 
\begin{equation}
 N(y')=\frac{4 \pi}{g} \frac{F(y')}{\Omega}
 \label{CD_1}
\end{equation}
where $g$ is the fluorescence efficiency in ergs per molecule per second, and $\Omega$ is the solid angle subtended by the aperture in steradians. $\Omega$ is calculated as the product of the slit width and size of the aperture used to extract the spectrum. The value of $y$ in the model equation \ref{CD} is set equal to $y'$ in order to reproduce the observed data.

\subsection{Calculation of $Af\rho$}
$Af\rho$ is a directly measurable quantity which was first introduced by \citet{A'hearn_1984} based on a simple radial outflow model. It is a proxy for the amount of dust that is being produced. It represents the height of the cylinder of radius which is equivalent to the projected aperture. It is calculated as 
\begin{equation}
Af\rho=\frac{(2 \Delta r)^{2}}{\rho} \frac{F_{comet}}{F_{sun}} \frac{1}{S(\theta)}                                                                                                                                                                                                                                                                                                                                                                                                                                             \end{equation}
where $\Delta$ and $\rho$ are in kilometer and $r$ is in AU. $F_{sun}$ is the solar flux at 1 AU and $F_{comet}$ is the observed cometary flux. This quantity is independent of the aperture used, and represents the amount of dust within the cylinder. $S(\theta)$ is called as the phase function\footnotemark[4]. This is incorporated to minimize the effect of varying scattering angle. \citet{HB} have discussed the specifications of the HB narrow-band filters, their continuum points and have defined the flux standards for the same. The blue, green and red continuum points are defined at 4450, 5260 and 7128 Angstroms respectively with a bandwidth of 60 Angstroms. The transmission profiles of these filters were verified in our laboratory using a UV-Visible spectrometer and these profiles were used to extract the continuum point fluxes from the comet's spectra. These fluxes were then used to calculate the $Af\rho$ values in the following way. 
\begin{figure*}
  \centering
  \includegraphics[width=1.1\textwidth]{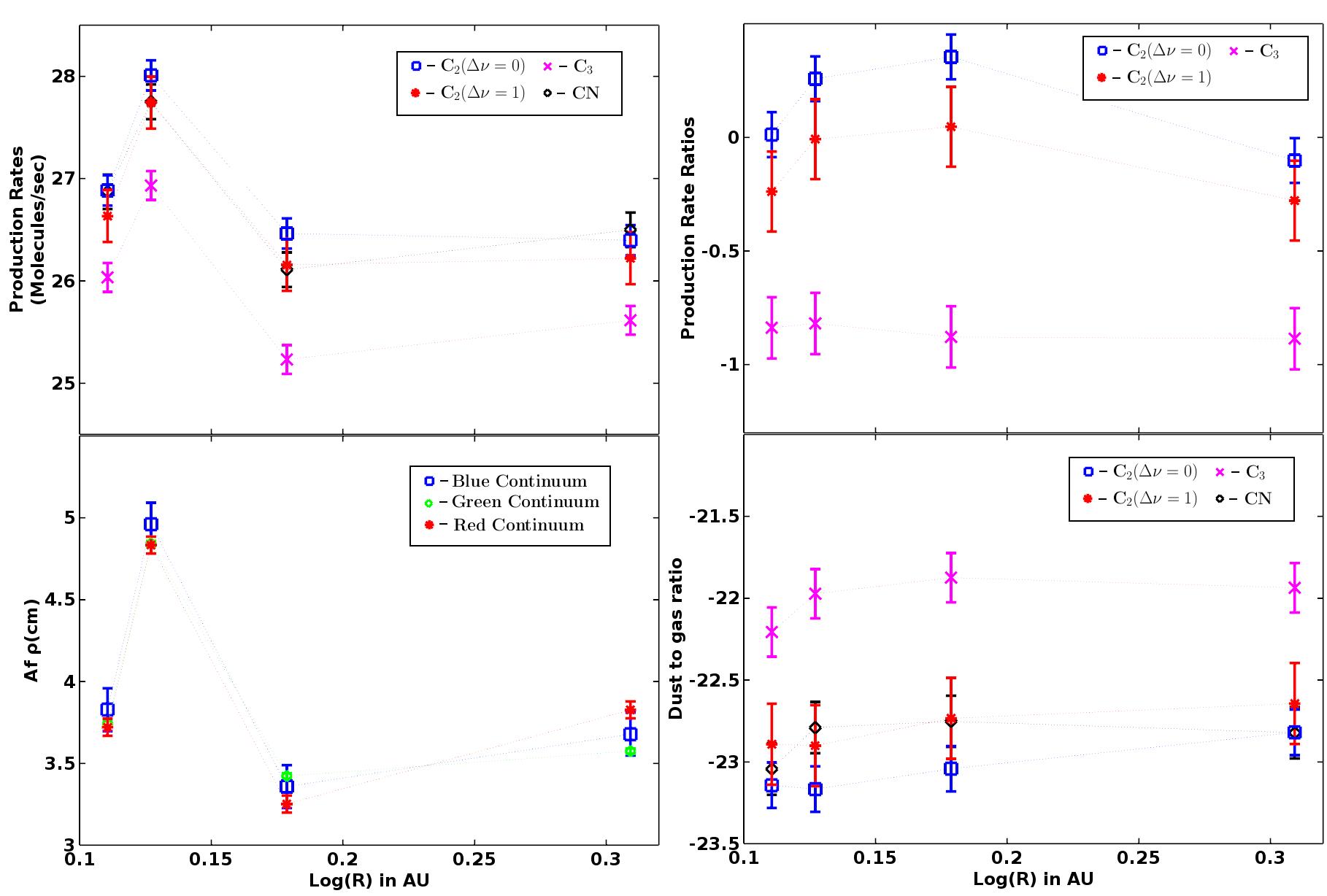}
  \caption{Top Left - Variation of production rate with heliocentric distance; Top Right - Variation of production rate ratios with heliocentric distance; Bottom Left - Variation of Af$\rho$ with heliocentric distance; Bottom Right - Variation of dust to gas ratio with heliocentric distance }
  \label{variation}
\end{figure*}
The transmission profiles of the HB filters were re-sampled to match the resolution of the comet spectrum and their product was calculated. The area under the curve of this product gave the cometary flux in the respective continuum band. The magnitude values of the solar analog stars from \citet{HB} at the continuum points recalculated for a distance of 1 AU were used to obtain the solar flux values.\footnotetext[4]{From \url{http://asteroid.lowell.edu/comet/dustphase.html} by D.Schleicher in May 2010} \label{afrho} 

\subsection*{Error Calculations}
The statistical error values obtained as the standard deviation of the multiple observations taken in the month of May, when the signal to noise ratio was worst, are considered as the upper limit for the errors at all the other epochs. \label{errors}

\section{Results and Discussion}
We have followed the comet over a period during which its heliocentric distance varied from 1.29 AU, just prior to perihelion to around 2.05 AU post perihelion. Various molecular emission lines like C$_{2}$, C$_{3}$, CN, NH$_{2}$, CH, O[I] were clearly seen in the comet spectrum throughout this range. The most prominent of them being the C$_{2}$ molecule, which was quite dominant throughout the time that we have followed the comet. Apart from the C$_{2}$ emission band, those of CN and C$_{3}$ were also quite prominent.

\subsection{Production Rates and Ratios}

The production rates were calculated for these molecules using equation \ref{Q_eq}. The production rates and ratios calculated for the various molecular band emissions at different heliocentric distances are given in table \ref{Q}. The ratios are taken with respect to CN molecule. The values in the parenthesis represent the errors in logarithmic scale. The variation of production rate and the ratios with heliocentric distance is shown in figure \ref{variation}. The values given in the Y axis are in log scales. \\

According to \citet{b1}, the variations in the gas ratios as well as dust to gas ratios are quite negligible unless the heliocentric distance varies more than a factor of 2. The production rate ratios do not show any particular trend with the heliocentric distance. \citet{b1} and \citet{Cochran2012} have set different limits for carbon depleted class of comets. The former defines the carbon depleted class as those for which log(Q(C$_{2}$)/Q(CN)) $<$ -0.18 whereas the latter defines this class as those comets for which log(Q(C$_{2}$)/Q(CN)) $<$ 0.02 and log(Q(C$_{3}$)/Q(CN)) $<$ -0.86. The daughter scale lengths that we have used are taken from \citet{b1}, whereas they differ from what \citet{Cochran2012} have used. So a direct comparison with \citet{Cochran2012} is not possible. If we take the mean for the ratios from our observations, we find that the C$_{2}$ to CN ratio seems to be within the class of typical comets. 
\begin{table*}
\caption{Gas Production Rates (molecules/sec in Log scale) at different heliocentric distances}
\begin{tabular}{c | c  | c | c | c | c |c| c | c|c|c|c|}
\hline \\
     &      &          &            &   \multicolumn{4}{c}{Production Rate(error)} & &\multicolumn{3}{c}{Production Rate Ratio with respect to CN(error)} \\ 
 \cline{5-8} \cline{10-12} \\
Date & R$_{h}$   & $\Delta$ & log $\rho$ & CN & C$_{2}$ & C$_{2}$ & C$_{3}$ & & C$_{2}$ & C$_{2}$ & C$_{3}$ \\ 
     & (AU) & (AU) & (km) & & $\Delta \nu = 0$ & $\Delta \nu = 1$ & & & $\Delta \nu = 0$ & $\Delta \nu = 1$ &  \\
\hline 
27/01/2015 & 1.29 & 0.692 & 4.60 & 26.87(0.17) & 26.89(0.15) & 26.63(0.25) & 26.03(0.14) & & 0.02(0.10) & -0.24(0.18) & -0.84(0.13) \\
\hline
23/02/2015 & 1.34 & 1.206 & 4.84 & 27.75(0.17) & 28.01(0.15) & 27.74(0.25) & 26.93(0.14) & &0.26(0.10) & -0.01(0.18) & -0.82(0.13) \\ 
\hline
23/03/2015 & 1.50 & 1.665 & 4.45 & 26.11(0.17) & 26.47(0.15) & 26.16(0.25) & 25.23(0.14) & &0.36(0.10) & 0.05(0.18) & -0.88(0.13) \\
\hline
19/05/2015$^*$ & 2.03 & 2.22 & 4.34 & 26.50(0.17) & 26.40(0.15) & 26.22(0.25) & 25.61(0.14) & &-0.10(0.10) & -0.28(0.18) & -0.89(0.13) \\
\hline
\end{tabular}
\hspace*{-1in}$^*$  \small{The values mentioned for this epoch are the mean of 3 days of observations(18/05, 19/05 and 21/05/2015)}
\label{Q}
\end{table*}

\subsection{Production Rates from the Haser Model fits}

\begin{figure}
  \centering
  \includegraphics[width=0.5\textwidth]{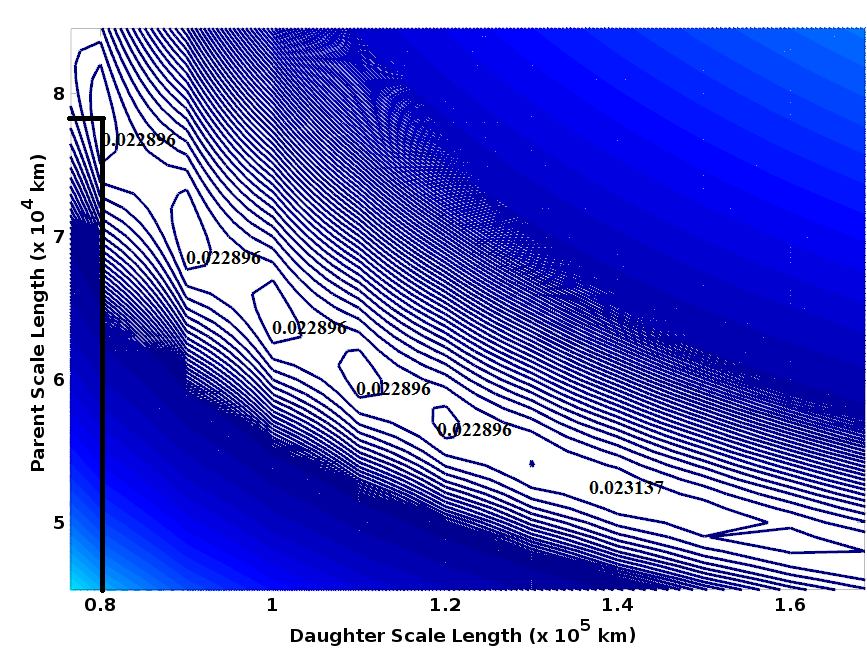}
  \caption{Contours showing the range of parent and daughter scale lengths that fit the observed data for C$_{2}$ molecule. The black lines indicate the value at which minimum $\chi^{2}$ was obtained.}
  \label{C2_contour}
\end{figure}

\begin{figure}  
  \includegraphics[width=0.5\textwidth]{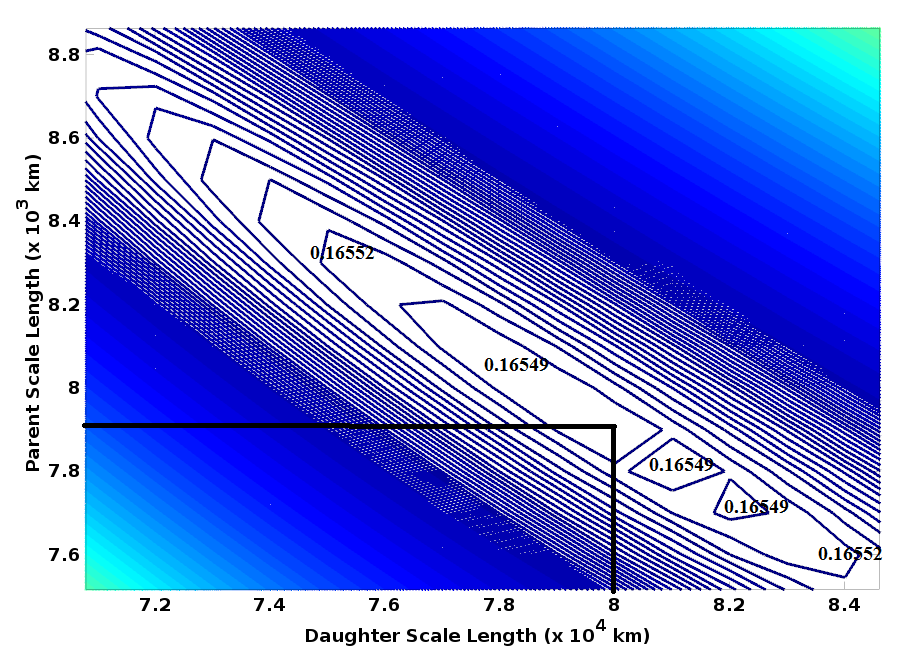}
\caption{Contours showing the range of parent and daughter scale lengths that fit the observed data for C$_{3}$ molecule. The black lines indicate the value at which minimum $\chi^{2}$ was obtained.} 
  \label{C3_contour}
\end{figure}

\begin{figure}
  \includegraphics[width=0.5\textwidth]{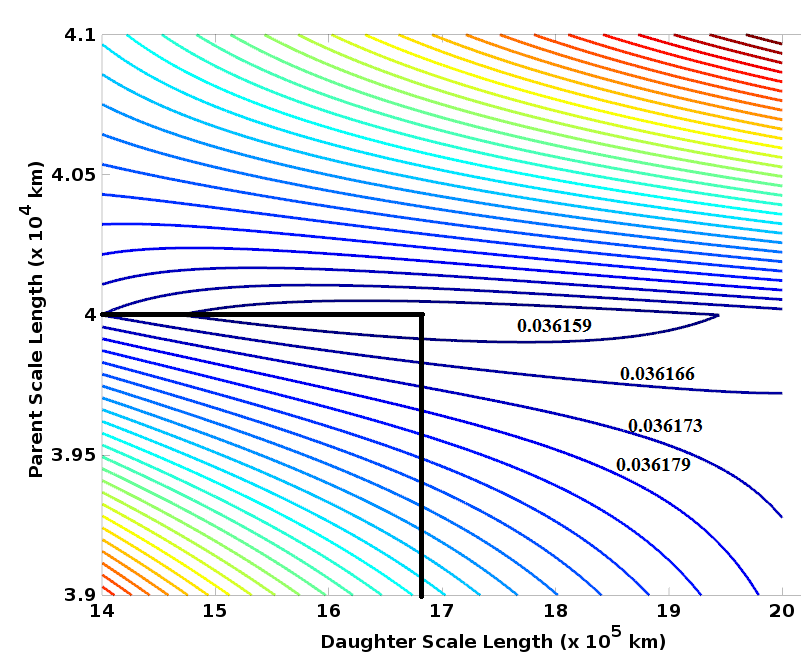}
  \caption{Contours showing the range of parent and daughter scale lengths that fit the observed data for CN molecule. The black lines indicate the value at which minimum $\chi^{2}$ was obtained.}
  \label{CN_contour}
\end{figure}

The production rates were also obtained by estimating the Haser model scale lengths of the observed molecular species. The observed column density as a function of projected distance from the nucleus for C$_{2}$, C$_{3}$ and CN were fitted with the theoretical values calculated from the Haser model. There are three unknown parameters : Production rate (Q), the parent and daughter scale lengths ($\beta_{0}$ and $\beta_{1}$ respectively). Q, $\beta_{0}$ and $\beta_{1}$ of the molecular species are to be determined while fitting the Haser model as described in section \ref{Haser}. The observed and theoretical column density, as obtained from equations \ref{CD} and \ref{CD_1} were normalized to one at the nearest point from the nuclear center. Normalization removes the production rate parameter from the equation and then leaves us with two undetermined parameters. The parent and daughter scale lengths were varied simultaneously and the column densities were calculated. The best fit value of the scale lengths were determined by minimizing the goodness parameter $\chi^{2}$  
\begin{equation}
\chi^{2}=\mathlarger{\mathlarger{\sum}}_{i} \frac{ (N_{o}^{i}-N_{h}^{i})^{2}}{N_{h}^{i}} 
\end{equation}
where $N_{o}^{i}$ is the observed column density and $N_{h}^{i}$ is the column density calculated from the Haser model for the i$^{th}$ aperture. Different authors have quoted various values of the scale lengths for the same species. Taking inspiration from the work of \citet{Langland} we have obtained our best fit scale lengths without referring to any of the literature value. This was done for the observational data of 23/02/2015, when the comet's heliocentric distance was 1.34 AU. Due to guiding issues and lower signal to noise ratios, the scale lengths could not be obtained at other epochs. There is no unique solution for the Haser model scale lengths. Instead, we get a family of solutions which fit the observed data. These are represented as contours. The contours for C$_{2}$, C$_{3}$ and CN molecular bands are shown in figures \ref{C2_contour}, \ref{C3_contour}, \ref{CN_contour}.
  
\begin{figure}
  \centering
  \includegraphics[width=0.5\textwidth]{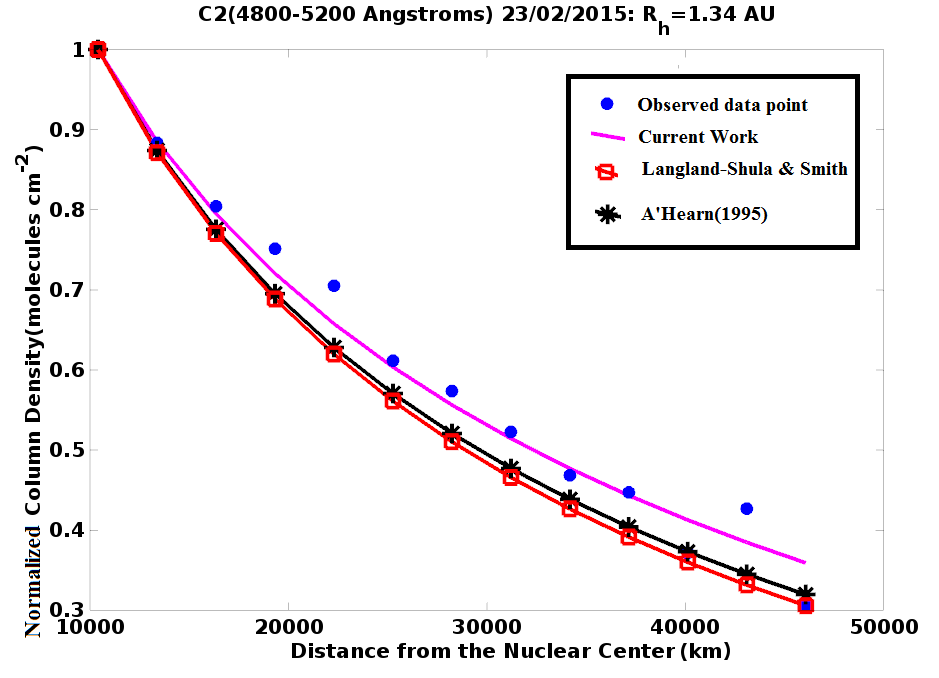}
  \caption{The observed column densities of C$_{2}$ as a function of the projected distance $r_{n}$ (in km) from the nucleus. The column densities calculated from Haser model using the current best fit scale lengths and scale lengths from \citet{b1} and \citet{Langland} are over-plotted}
  \label{C2_profile}
\end{figure}

\begin{figure}  
 \includegraphics[width=0.5\textwidth]{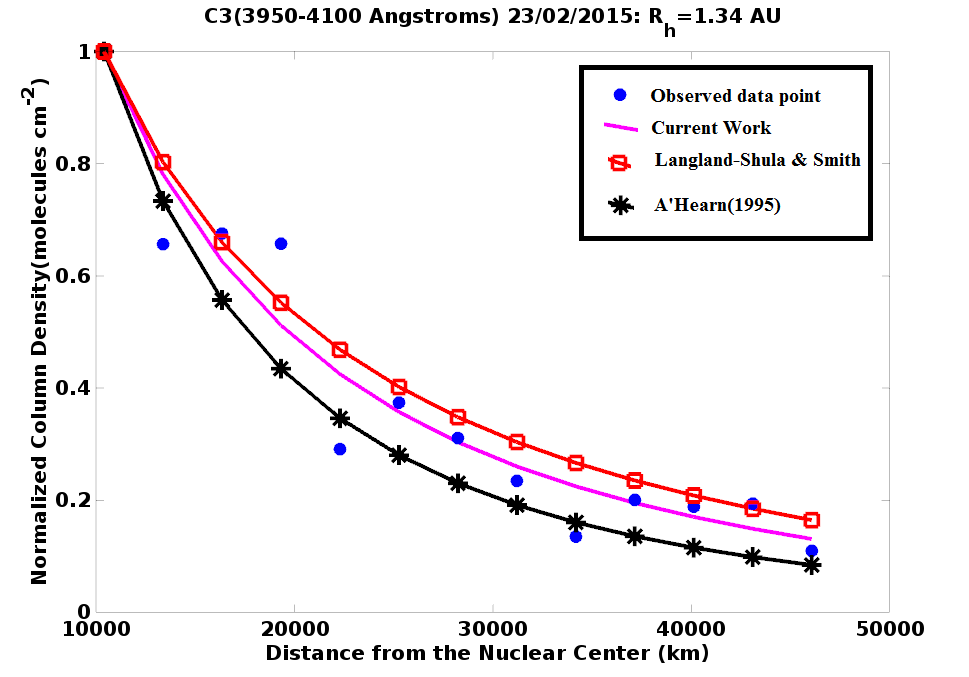}  
\caption{The observed column densities of C$_{3}$ as a function of the projected distance $r_{n}$ (in km) from the nucleus. The column densities calculated from Haser model using the current best fit scale lengths and scale lengths from \citet{b1} and \citet{Langland} are over-plotted} 
  \label{C3_profile}
\end{figure}

\begin{figure}
  \includegraphics[width=0.5\textwidth]{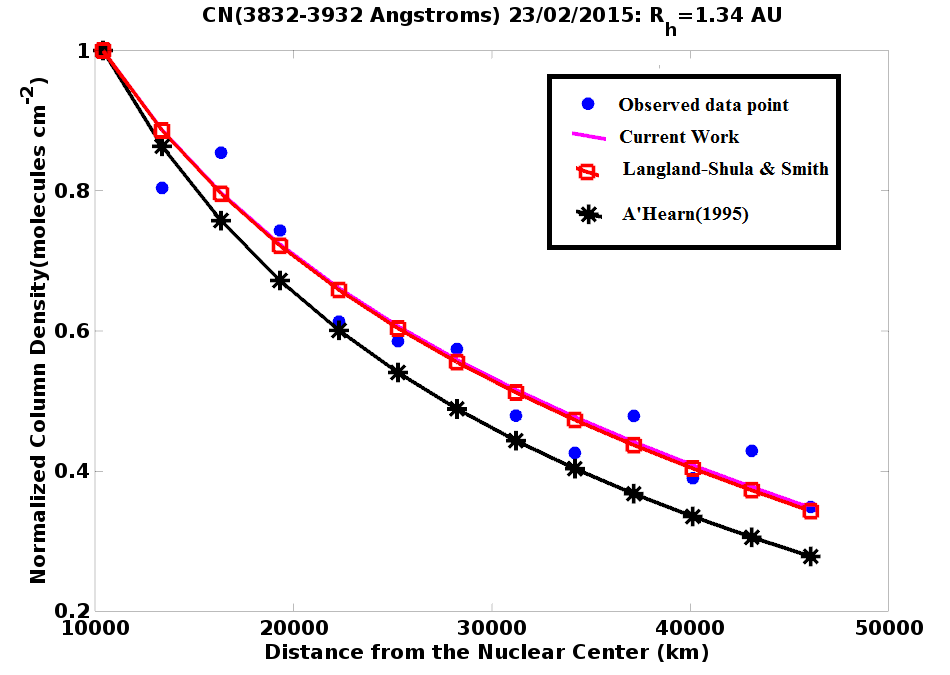}
  \caption{The observed column densities of CN as a function of the projected distance $r_{n}$ (in km) from the nucleus. The column densities calculated from Haser model using the current best fit scale lengths and scale lengths from \citet{b1} and \citet{Langland} are over-plotted}
  \label{CN_profile}
\end{figure}

There are many small contours, which represent multiple solutions. From the contours encompassing the smaller ones, we have obtained the range of parent and daughter scale lengths. For the C$_{2}$ molecule, the daughter scale length varies from 0.69 $\times$ $10^{5}$ km to 1.5 $\times$ $10^{5}$ km and parent scale length from 4.9 $\times$ $10^{4}$ km to 8.3 $\times$ $10^{4}$ km. For the C$_{3}$ molecule, the daughter scale length varies from 7.2 $\times$ $10^{4}$ km to 8.8 $\times$ $10^{4}$ km and parent scale length from 7.3 $\times$ $10^{3}$ km to 8.7 $\times$ $10^{3}$ km. In case of CN, although we get a minimum, it might not be the true value because of various uncertainties in the blue side of the spectrum. This might be due to uncertainties in the extinction correction. The column densities for Haser model scale lengths of all the molecular species from this work, \citet{b1} and \citet{Langland} are over plotted for comparison. These are shown in figures \ref{C2_profile}, \ref{C3_profile} and \ref{CN_profile}. The new scale lengths fit the observed data much better than the others, except for CN, where our scale lengths fit equally well as that of \citet{Langland}. The CN scale lengths are not as well constrained to a limited range and the derived numbers appear to be on the larger side.  The CN band is in the blue end of the spectrum where the atmospheric extinction is quite uncertain. Also the Haser model being a very simple model, may not be sufficient at explaining the observed profiles. \\
Once the scale lengths are obtained, the production rate is calculated using equation \ref{CD} for different distances from the center. The mean of the production rate value is taken as final estimated production rate of the molecular species. For comparison, we have also calculated the production rates using scale lengths from \citet{b1} and \citet{Langland}. The scale lengths and production rates obtained from the Haser model fitting are tabulated in table \ref{HM}. In order to see the trends in the production rates, the production rate ratios and the dust to gas ratios with heliocentric distance, we use the production rate values calculated from equation \ref{Q_eq} using a single daughter scale length.

\begin{table*}
\caption{Haser Model Scale lengths and production rates}
\begin{tabular}{c | c | c  | c | c |c | c | c | c | c | }
\hline \\
     &  \multicolumn{3}{c}{\citet{b1}} &\multicolumn{3}{c}{\citet{Langland}} &\multicolumn{3}{c}{Current Work}\\ 
 \cline{2-4} \cline{5-7} \cline{8-10} \\
  & $\beta_{0}$ & $\beta_{1}$ & Q & $\beta_{0}$ & $\beta_{1}$ & Q & $\beta_{0}$ & $\beta_{1}$ & Q \\ 
     & ($\times$ $10^{4}$ km) & ($\times$ $10^{4}$ km) & (mol/sec) & ($\times$ $10^{4}$ km) & ($\times$ $10^{4}$ km) & (mol/sec) & ($\times$ $10^{4}$ km) & ($\times$ $10^{4}$ km) & (mol/sec) \\
\hline 
C$_{2}$ & 3.95 & 11.85 & 27.73 & 5.4 & 6.5 & 27.71 & 7.8 & 8.0 & 27.96 \\

\hline

C$_{3}$ & 0.50 & 4.84 & 26.98 & 0.93 & 13.0 & 26.78 & 0.79 & 8.0 & 26.81 \\ 

\hline
CN  & 2.33 & 37.7 & 27.82 & 4.65 & 30.8 & 27.97 & 4.0 & 167.0 & 27.90 \\
\hline

\end{tabular}
\label{HM}
\end{table*}

\subsection{Dust Production : $Af\rho$ }

The $Af\rho$ was calculated for the blue continuum(4390-4510$\AA$), green continuum(5200-5320$\AA$) and the red continuum(7060-7180$\AA$) points of the HB filter set, as explained in the section \ref{afrho}. The log($Af\rho$) values are tabulated in the table \ref{afrho_table}. The values in the parenthesis represent the errors in logarithmic scale. It can be clearly noted from these values that within the errors, the scattering from the dust in the bluer wavelength side is greater than that of the red continuum. This indicates that there might be a large number of smaller dust particles. But it is the other way round at the last epoch at a heliocentric distance of 2.05 AU. This indicates that the comet was populated by smaller dust particles near perihelion which got depleted as the comet moved away. 
The dust to gas ratio shown in figure \ref{variation} is calculated as the ratio of Af$\rho$ to the production rates of the molecular gas species like CN, C$_{2}$ and C$_{3}$. This remains almost constant within the error limits.\citet{b1} has shown the variation of dust to gas ratio as a function of the perihelion distance of the comet. The values obtained for this comet are consistent with this trend.

\begin{table}
 \caption{$Af\rho$(cm in Log scale) for different continuum bands}
 \begin{tabular}{c|l|c|l|}
 \hline
 Date & Blue           & Green           & Red  \\
      & Continuum      & Continuum       & Continuum \\
 \hline
 
 27/01/2015 & 3.83(0.13) &  3.74(0.02) & 3.72(0.05) \\

23/02/2015 & 4.96(0.13) & 4.84(0.02) & 4.83(0.05) \\

23/03/2015 & 3.36(0.13) & 3.42(0.02) & 3.25(0.05) \\ 

19/05/2015$^*$ & 3.68(0.13) & 3.58(0.02) & 3.83(0.05) \\
 \hline
  \end{tabular}
  \label{afrho_table}
\hspace*{0.2in}$^*$  \small{The values mentioned for this epoch is the mean of 3 days of observations(18/05, 19/05 and 21/05/2015)}
\end{table}

The gas production rate as well as $Af\rho$ increase after perihelion and show a decreasing trend only after February 2015. It is interesting to note the simultaneous increase in gas and dust which indicates an increase in the overall activity of the comet after its perihelion passage. This kind of asymmetry has been seen in many comets. In fact \citet{b1} have tabulated the difference in the pre and post perihelion values for different class of comets with positive values indicating post perihelion activity. Although we do not have data points at exactly the same distance for pre and post perihelion passages, we can perhaps say that this comet may have a large positive asymmetry. This indicates that there might be volatile material present beneath the surface of the comet or the surface of the nucleus consists of layers of ice that have different vaporization rates.
\section{Conclusions}
We have carried out spectral study of the comet C/2014 Q2 (Lovejoy) which shows strong emission lines of C$_{2}$, C$_{3}$ and CN. The molecular production rates and the quantity Af$\rho$ are estimated. We arrive at the following conclusions:
\begin{itemize}
 \item The comet shows a large positive asymmetry, which indicates increase in the post-perihelion activity.
 \item Considering the limits set by \citet{b1} for the carbon abundance, C/2014 Q2 is found to be in the typical class of comets.
 \item The dust to gas ratio remains fairly constant within the observed heliocentric range which is in agreement with \citet{b1} 
 \item The $Af\rho$ values indicate that the dust distribution favours smaller dust particles near perihelion which depletes as the comet moves away.
 \item New Haser model scale lengths were obtained which best fit the observed data. These differ significantly from the previously quoted values. This might be due to the comet's intrinsic nature itself or affected by the solar activity at the time of observation.
\end{itemize}
A more exhaustive study is required to confirm some of the conclusions drawn here.

\section*{Acknowledgments}
This work is supported by the Dept of Space, Govt of India. We acknowledge the local staff at the Mount Abu Infra-Red Observatory for their help. We would also like to thank Prof. David Schleicher for his invaluable inputs and suggestions. We acknowledge with thanks Prof Jeremy Tatum for useful discussions. We would like to specially thank Ms. Navpreet Kaur and Mr. Sameer for their help in the observations. We also thank our colleagues in the Astronomy \& Astrophysics division at PRL for their comments and suggestions.

\label{lastpage}

\end{document}